\RequirePackage{lineno} 
\documentclass{elsart} 
  \usepackage{epsfig,graphics,psfrag,amsmath}

 \usepackage{cases}
 \usepackage{txfonts}
 \usepackage{mathrsfs}

 \usepackage{amssymb}
 \usepackage{dcolumn}
 \usepackage{bm}
 \usepackage{subfigure}
 \usepackage{floatflt}
 \usepackage{float}
 \usepackage{rotating}
 \usepackage{multirow}
 \usepackage[bookmarksnumbered,bookmarksopen,colorlinks,citecolor=blue,linkcolor=blue]{hyperref} %
 \usepackage{booktabs}

 \usepackage{sidecap}
  \usepackage{lineno}

\def\esym{$E_{\rm sym}(\rho)$}

\def \rthe{$R({\rm t/^3He})$}

\begin{document}
 \bibliographystyle{unsrt}

\begin{frontmatter}

\title{Revisit to  the yield ratio of triton and $^3$He  as an indicator of  neutron-rich neck emission}


\author[THU]{Yijie Wang\thanksref{info_wang}},
\author[gxnu,gxkl]{Mengting Wan},
\author[THU]{Xinyue Diao},
\author[THU]{Sheng Xiao},
\author[THU]{Yuhao Qin},
\author[THU]{Zhi Qin},
\author[THU]{Dong Guo},
\author[THU]{Dawei Si},
\author[THU]{Boyuan Zhang},
\author[THU]{Baiting Tian},
\author[THU]{Fenhai Guan},
\author[THU]{Qianghua Wu},
\author[IMP]{Xianglun Wei},
\author[IMP]{Herun Yang},
\author[IMP]{Peng Ma},
\author[IMP]{Rongjiang Hu},
\author[IMP]{Limin Duan},
\author[IMP]{Fangfang Duan},
\author[IMP]{Junbing Ma},
\author[IMP]{Shiwei Xu},
\author[IMP]{Qiang Hu},
\author[IMP]{Zhen Bai},
\author[IMP]{Yanyun Yang},
\author[HZU,IMP]{Jiansong Wang},
\author[HNU]{Wenbo Liu},
\author[HNU]{Wanqing Su},
\author[HNU]{Xiaobao Wei},
\author[HNU]{Chunwang Ma},
\author[southchina,SINAP]{Xinxiang Li},
\author[SINAP,SARI]{Hongwei Wang},
\author[ciae]{Yingxun Zhang},
\author[UMCS]{Micha{\l} Warda},
\author[UMCS]{Arthur Dobrowolski},
\author[UMCS]{Bo{\.z}ena Nerlo-Pomorska},
\author[UMCS]{Krzysztof Pomorski},
\author[gxnu,gxkl]{Li Ou\thanksref{info_ou}},
\author[THU,cicq]{Zhigang Xiao\thanksref{info_xiao}}

\address[THU]{Department of Physics, Tsinghua University, Beijing 100084, China;}
\address[gxnu]{College of Physics and Technology, Guangxi Normal University, Guilin 541004, China;}
\address[gxkl]{Guangxi Key Laboratory of Nuclear Physics and Technology, Guangxi Normal University, Guilin 541004, China;}
\address[UCAS]{University of Chinese Academy of Sciences, Beijing 100049, China;}
\address[IMP]{Institute of Modern Physics, Chinese Academy of Sciences, Lanzhou 730000, China;}
\address[HZU]{School of Science, Huzhou University, Huzhou, 313000, China;}
\address[BNU]{College of Nuclear Science and Technology, Beijing Normal University, Beijing 100875, China;}
\address[HNU]{Institute of Particle and Nuclear Physics, Henan Normal University, Xinxiang 453007, China;}
\address[SINAP]{Shanghai Institute of Applied Physics, Chinese Academy of Science, Shanghai 201800, China;}
\address[SARI]{Shanghai Advanced Research Institute, Chinese Academy of Science, Shanghai 201210, China;}
\address[southchina]{School of Nuclear Science and Technology, University of South China, Hengyang 421001, China;}
\address[ciae]{China Institute of Atomic Energy, Beijing 102413, China;}
\address[UMCS]{Institute of Physics, Maria Curie Sk{\l}odowska University, 20-031 Lublin, Poland}
\address[cicq]{Collaborative Innovation Center of Quantum Matter, Tsinghua University, Beijing 100084, China;}

\thanks[info_wang]{E-mail:~yj-wang15@tsinghua.org.cn}
\thanks[info_ou]{E-mail:~liou@gxnu.edu.cn}
\thanks[info_xiao]{E-mail:~xiaozg@tsinghua.edu.cn}

\end{frontmatter}

\begin{frontmatter}
 \date{\today}
 \begin{abstract}
The  neutron rich neck zone created in heavy ion reaction is experimentally probed by the production of the $A=3$ isobars. The energy spectra and angular distributions of triton and $^3$He are measured with the CSHINE detector in $^{86}$Kr +$^{208}$Pb reactions at 25 MeV/u. While the energy spectrum of $^{3}$He is harder than that of triton, known as "$^{3}$He-puzzle",  the yield ratio $R({\rm t/^3He})$ presents a robust rising trend with the polar angle in laboratory. Using the fission fragments to reconstruct the fission plane, the enhancement of  out-plane  $R({\rm t/^3He})$ is confirmed in comparison to the in-plane ratios. Transport model simulations reproduce qualitatively the experimental trends, but the quantitative agreement is not achieved. The results demonstrate that a neutron rich neck zone is formed in the reactions. Further studies are called for to understand the clustering and the isospin dynamics related to neck formation.   

 \end{abstract}
\end{frontmatter}


 \maketitle


  
   \section {Introduction  } 
   Heavy ion reaction (HIR) is a micro-laboratory for investigating the properties of the nuclear equation of state (nEoS), particularly the nuclear symmetry energy \esym ~\cite{Li:2021thg,Huth:2021bsp,Steiner:2004fi,Oertel:2016bki,Li:2014oda,Li:2008gp}.  The accurate constraint of \esym~ is crucial for both nuclear- and astro-physics, and becomes unprecedentedly important since the detection of the gravitational wave from the neutron star merging event GW170817 \cite{LIGOScientific:2017vwq,LIGOScientific:2018cki,De:2018uhw}.  
  Although great progress has been made via the detection of isobaric yield ratios in HIRs, like n/p \cite{Li:1997rc}, $\rm t/^3He $ \cite{Zhang:2005sm,Chen:2003qj},  $\pi^-/\pi^+$ \cite{Li:2002qx,Xiao:2008vm,SRIT:2021gcy,Li:2005kqa}, $K^0/K^+$ \cite{Ferini:2006je} and $\Xi^-/\Xi^0$ \cite{Yong:2022pyb}, the transport model evaluation project (TMEP) has been launched to benchmark the method to allocate the uncertainties in model analysis of \esym~  \cite{Xu:2019hqg,Wang:2020vwb,Zhang:2017esm,Ono:2019ndq,Colonna:2021xuh}, and the efforts are ongoing to search  novel probes to explore the effects of \esym~ in HIRs \cite{Zhang:2017xtk,Ou:2015jan,Wang:2021mrv,Wang:2022ysq}.

  The determination of \esym~ relies on the comprehensive view  of the underlining dynamic process of HIRs, particularly that involving the isospin degree of freedom (IDOF) \cite{Hudan:2014xki}. During the thermal and chemical evolution of the reaction,  a low-density neutron rich neck zone is formed as the result of the complicated dynamics \cite{Poggi:2001wkr,di2006neck}.
  The neck zone has been explored to understand the mechanism of intermediate mass fragment formation \cite{Toke:1995zz,Dempsey:1996zz,Ramakrishnan:1998zz,Hudan:2005au}, isotopic cluster emission  \cite{Sobotka:1997zz,Laforest:1999zz,zhang2015correlation,Feng:2016hqa} and neutron-proton equilibration \cite{RodriguezManso:2017emd,Wang:2018bma,Bougault:2017xtb,May:2018ulz,INDRA:2022amz}. Because of the density gradient and the isospin migration, the neck zone provides a beneficial environment to study the \esym~ \cite{Bougault:2017xtb,INDRA:2022amz}.
  For more discussions about neck zone, one can refer to the review articles of heavy ion reactions from the experimental  \cite{Poggi:2001wkr,di2006neck,DeFilippo:2013ipa} and  theoretic points of view \cite{Baran:2001pz,Baran:2003mq,Baran:2005ce,Colonna:2013jqa,Colonna:2020euy}.

  Among the probes using the light charged particles (LCPs), the yield ratio of  $\rm t/^3He $, written as \rthe, has been particularly identified to probe the enriched feature of isospin dynamics in HIRs. Transport model calculations demonstrate that the \rthe~ at intermediate energies HIRs depends on the stiffness of \esym~ \cite{Chen:2003qj,Chen:2003ava}. 
 At high energy HIRs,  \rthe~ depends more sensitively on the value of \esym~  \cite{Wang:2014aba} and the specific form of the interaction potential \cite{Li:2005kqa,Gaitanos:2004zh}, but less dependent on the slope of \esym~ \cite{Yong:2009te}. In addition, \rthe~  reflects the isospin dependent nucleon density in the reactions \cite{Chomaz:1998tp,Dempsey:1996zz,Albergo:1985zz}.  Experimentally, the yield ratios of various mirror nucleus pairs, including  the \rthe, led to the discovery of isospin fractionation \cite{Xu:1999bs}. It has been suggested that  more neutron-rich particles are emitted at midrapidity, as inferred by the \rthe, which correlates positively with the charge number of projectile-like fragments\cite{Dempsey:1996zz} but reversely with the center of mass energy \cite{Famiano:2006rb}. Similarly, in high-energies HIR, the \rthe~ reflects the neutron enrichment of the emission source\cite{Nagamiya:1981sd,Dempsey:1996zz,Veselsky:2000kp} and isospin mixing during the collision \cite{FOPI:1999gxl}. Recently, the \rthe~ has also been used to study the pick-up mechanism of pre-equilibrium light nucleus production in the pion scattering experiment \cite{Gurov:2014hea}.
  
  
  Despite of the enormous progress of the studies on the triton and $^3$He emission, some questions remain unclear and require further studies. For example, when considering the spectra of $^3$He, there is an anomalous phenomenon that the yield of high energy $^3$He is relatively larger, compared to that of triton  \cite{Poggi:1995zna,bougault1997possible,Liu:2012kj,EOS:1994dsb,bonasera2000critical} or $^4$He \cite{Poggi:1995zna,Neubert:1999sv,Liu:2012kj,EOS:1994dsb,bonasera2000critical}. This phenomenon has been called “$^3$He-puzzle” \cite{Poggi:1995zna,bougault1997possible,bonasera2000critical}. While the energy spectra are suffering  “$^3$He-puzzle”, the yield ratio of t and $^3$He is sensitive to the neutron-to-proton ratio (N/Z) of the emission system \cite{Veselsky:2000kp,Shetty:2003ar,May:2018ulz,Piantelli:2021ybu}. The excitation function of \rthe~ measured by the FOPI collaboration \cite{FOPI:2010xrt} can not be reproduced with a single model \cite{Wang:2014aba}. More interestingly, the results of the INDRA experiment suggest that the t and $^3$He isobars seem to dominate the neutron enrichment of the neck zone \cite{INDRA:2022amz}.
  
  Due to the enriched but not-well-understood information carried by triton and  $^{3}$He  coupling to both the isospin transport and the neck formation in HIRs, we are motivated to explore the emission of these two isobars by looking at the energy spectra and the yield ratio $R({\rm t/^3He})$  over wide angular range, and to link the ratio to the feature of neck emissions.  In this letter, the energy spectra of  t and  $^{3}$He at different angles are measured in the reactions of $^{86}$Kr+$^{208}$Pb at 25 MeV/u. The  distributions of $R({\rm t/^3He})$  as a function of the laboratory polar angle and of the angle with respect to the fission plane are analyzed.  The comparison of the experimental data to the transport model simulation is discussed. The paper is organized as following. Section 2 and 3 present the experimental setup and the description of the transport model, respectively. Section 4 is the results and the discussions, and section 5 is the summary.

 \section  {Experimental setup } 
\begin{figure}[h] 
 \centering
 \includegraphics[angle=0,width=0.45\textwidth]{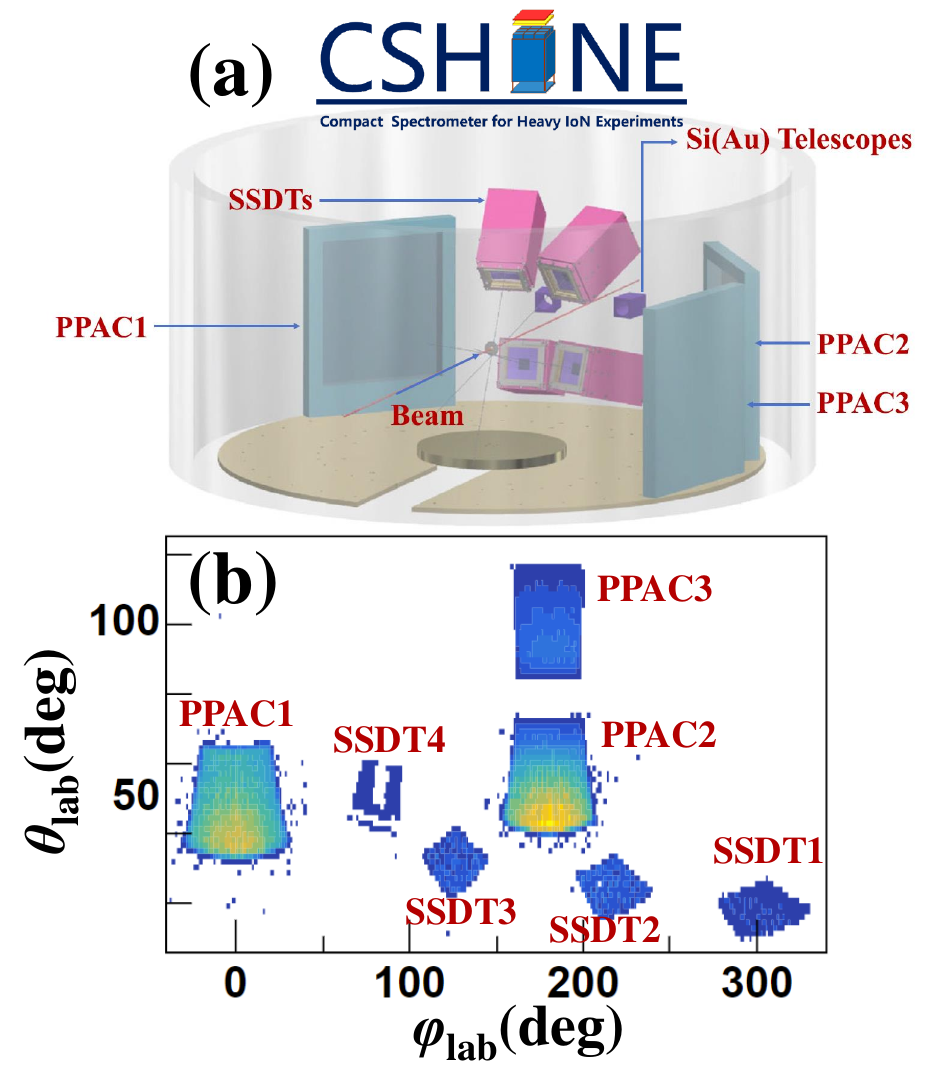}
 \caption{(Color online)  (a) The experimental setup of CSHINE. (b) The spatial  coverage of SSDTs and PPACs on $\theta-\varphi$ plane in laboratory reference frame.}
 \label{CSHINE}
 \end{figure}

The experiment was conducted at the Compact Spectrometer for Heavy IoN Experiment (CSHINE) \cite{Guan:2021tbi,Wang:2021jgu}, built at the final focal plane of the Radioactive Ion Beam Line at Lanzhou (RIBLL-I) \cite{SUN2003496}.  The  $^{86}$Kr beam of 25 MeV/u was extracted from the cyclotron of the Heavy Ion Research Facility at Lanzhou (HIRFL) \cite{XIA200211}, bombarding a natural lead target installed in the scattering chamber with the radius $R\approx 750$ mm. The target thickness is about $1~ {\rm mg/cm^2}$. The LCPs from the reactions were measured by  4 silicon-strip detector telescopes (SSDTs), covering  the angular range from  $10^\circ$ to $60^\circ$ in laboratory. 
Each SSDT consists of three layers, namely, one single-sided silicon-strip detector (SSSSD) for $\Delta E_{1}$ and one double-sided silicon strip detector (DSSSD) for $\Delta E_{2}$, backed by a $3\times 3 $ CsI(Tl) crystal hodoscope with the length of 50 mm for the energy deposit $E$. The granularity of the SSDT is  $4\times 4~ {\rm mm^2}$, giving about $1^\circ$ angular resolution.  The energy resolution of the SSDT is better than $2\%$, and the isotopes up to $Z=6$ can be identified. Multi hits and signal sharing are carefully treated in the track recognition, and the track recognition efficiency is about 90\% \cite{Guan:2021nfk}.  
In order to tag the reaction geometry, the fission fragments (FFs) were detected  by 3 parallel plate avalanche counters (PPACs), each of which had a sensitive area of $240\times 280~ {\rm mm^2}$. The perpendicular distance of the PPACs to the target is about 428 mm. The coverage of the PPACs ensures a high efficiency to measure the LCPs in coincidence with the FFs  selecting the semi-peripherial collisions. Fig.  \ref{CSHINE} presents the experimental setup (a) and the spatial coverage of the SSDTs and the PPACs (b).

  \section  {Theoretical Model}
 
 A hybrid model by the improved quantum molecular dynamics model  (ImQMD05) coupled with statistic decay afterburner (GEMINI) was used for theoretical simulation in this work.
 The ImQMD05 \cite{Zhang:2020dvn} was used to simulate the nucleon transport process in HIRs. And the GEMINI \cite{Charity:1988zz,Charity:2001yu} was followed to obtain the final state productions of the reactions. The ImQMD05 model is an improved version from the original quantum molecular dynamics code \cite{Aichelin:1989,Aichelin:1991}, and is widely used to understand the nuclear reaction dynamics by event-by-event analysis in both low and intermediate energies HIRs. The mean field part of the ImQMD05 model used here includes the symmetry potential energy part. And the local nuclear potential energy density functional in the ImQMD05 model is written as
 
 \begin{align}\label{Vloc}
	V_{\rm{loc}}&=\frac{\alpha}{2}\frac{\rho ^{2}}{\rho_{0}}+\frac{\beta }{\eta +1}%
	\frac{\rho ^{\eta +1}}{\rho_{0}^{\eta }}+\frac{g_{\rm{sur}}}{2\rho_{0}}\left(\nabla \rho \right)^{2}\\ \nonumber
	&+\frac{g_{\rm{sur,iso}}}{\rho_{0}}[\nabla(\rho_{\rm n}-\rho_{\rm p})]^{2} 
	+g_{\rho\tau}\frac{\rho^{8/3}}{\rho_{0}^{5/3}}+\frac{C_{\rm s}}{2} \frac{\rho^{\gamma+1}}{\rho_0^{\gamma}} \delta^{2},
\end{align}

where $\rho$, $\rho_{\rm n}$, $\rho_{\rm p}$ are the density of nucleon, neutron, and proton, respectively. $\delta = (\rho_{\rm n}-\rho_{\rm p})/(\rho_{\rm n}+\rho_{\rm p})$ is the isospin asymmetry degree. The parameters in Eq. \eqref{Vloc} except $C_{\rm s}$, which are listed in Table \ref{parameter}, are obtained directly from MSL0 parameter set \cite{Chen:2010qx}. $C_{\rm s}$ is determined by the symmetry potential energy at saturation density. Together with various $\gamma$, one can get the MSL0-like Skyrme interaction with various density dependent symmetry potential energy. The reaction was simulated with impact parameter in the range of  $1.0\le b \le 7.0$ fm by step of $\Delta b = 1.5$ fm. At the end of dynamics evolution in ImQMD, the minimum spanning tree (MST) algorithm \cite{Aichelin:1991,Zhang:2012} was used to recognize the free nucleons and fragments formed in the evolution. Next, the statistical decay of excited fragments was performed with GEMINI model. At last, the information of  final state  particles will be obtained.

 \begin{table*}[h]
 \caption{\label{parameter}Parameter set used in the ImQMD calculations.}
 \begin{tabular}{ccccccccc}
 \hline\hline
 $\alpha $ & $\beta $ & $\eta $ &$ g_{\rm{sur}}$ & $g _{\rm{sur,iso}}$ & $ g_{\rho\tau }$  & $C_{\rm{s}}$  & $\rho_{0}$  \\
  (MeV)    & (MeV)    &           & (MeV~fm$^{2}$) & (fm$^{2}$)          & (MeV)       & (MeV)             &  (fm$^{-3}$) \\ \hline
  -254     & 185      &  5/3       & 21.0           & -0.82               & 5.51             & 36.0  & 0.160 \\
 \hline\hline
 \end{tabular}
 \end{table*}

 \section  {Results and Discussions} 

\begin{figure}[h] 
 \centering
 \includegraphics[angle=0,width=0.85\textwidth]{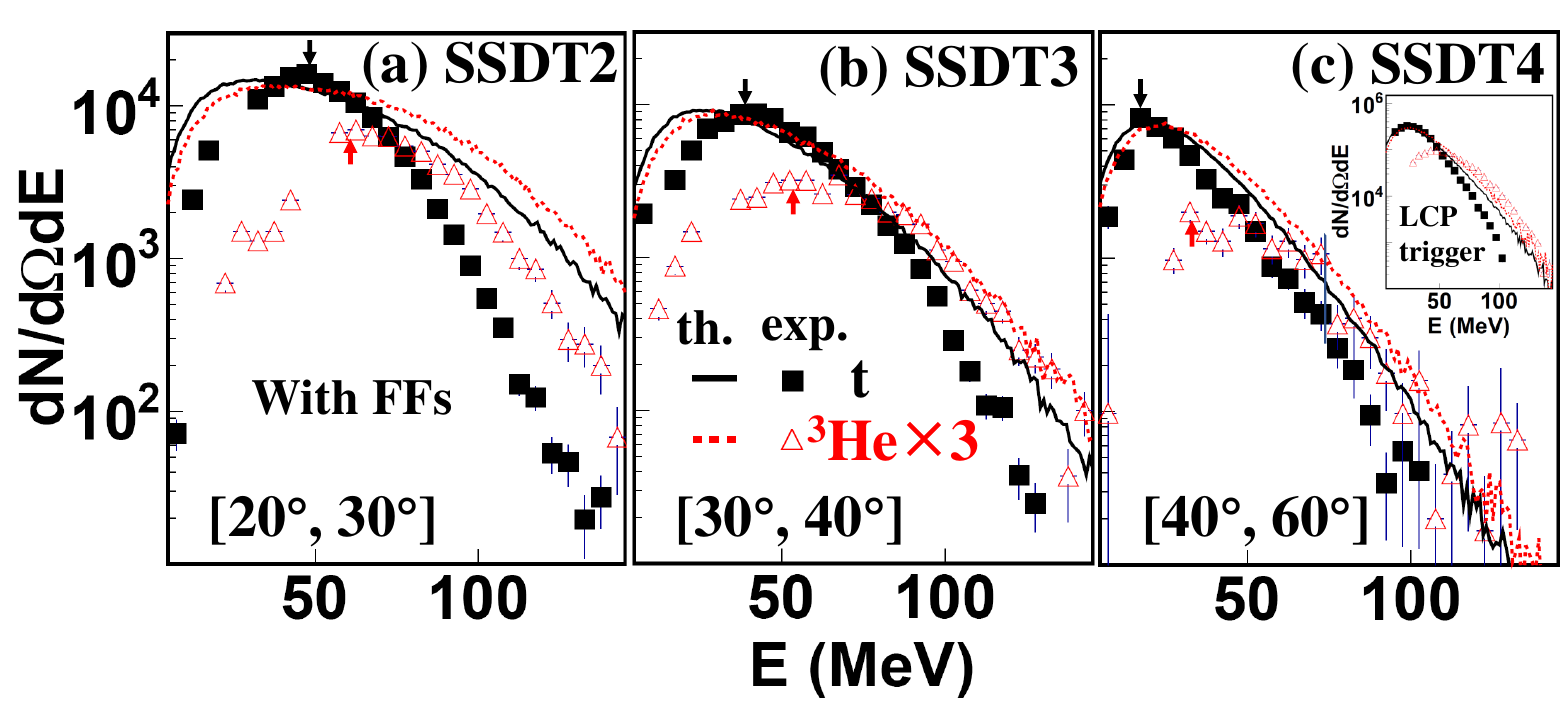}
 \caption{(Color online)  The energy spectra of  triton (solid) and $^3$He (open) in $20^\circ \le \theta_{\rm lab} \le 60^\circ$ covered by SSDT2 to SSDT4 in coincidence with the FFs. The arrows represent the peak position of each energy spectrum. Displayed in the inset of panel (c) is the spectra of t and $^3$He with trigger condition of $M_{\rm lcp}  \ge  2$.    }
 \label{energy_spectra}
 \end{figure}

We first analyze the energy spectra, which contain  thermal and dynamical information of the particle emission. Fig.  \ref{energy_spectra} presents the energy spectra of triton (solid squares) and $^3$He (open triangles)  in different angular ranges corresponding to SSDTs 2 to 4. To reduce the influence of quasi-projectiles, the data of SSDT1 covering $10-20^\circ$ in the laboratory is not counted here. It is shown that the spectrum of $^3$He is generally harder than that of triton, leading to a larger average kinetic energy of the former. The difference between triton and $^3$He is more pronounced at forward angles than at large angles. This observation of "$^{3}$He-puzzle" is in accordance with the previous measurements at high beam energies  \cite{Gutbrod:1976zzr,Poggi:1995zna,EOS:1994dsb,FOPI:1996cjz,Xi:1998zza,bonasera2000critical,Liu:2012kj,Raduta:2007qz,FOPI:2010xrt}. Consistently, our experimental results supports that the "$^{3}$He-puzzle" is more pronounced in the heavy system than that in light systems reported in the Fermi energy range  \cite{Xi:1998zza,Raduta:2007qz,Liu:2012kj}.

The “$^{3}$He-puzzle" has been interpreted by two possible scenarios: sequential decay \cite{bougault1997possible} and coalescence model \cite{Neubert:1999sv}. In the sequential decay scenario, the difference between $^{3}$He and t is influenced by the Coulomb barrier, for which $^{3}$He is emitted at an earlier stage with high temperature to overcome the Coulomb barrier higher than that of t \cite{bougault1997possible}.  In coalescence scenario, which was applied to interpret the difference between $^{3}$He and $\alpha$ particles \cite{Neubert:1999sv}, the former is dominantly produced by the coalescence of preequilibrium nucleons, delivering larger mean kinetic energy. These two explanations are qualitatively in agreement, suggesting that $^{3}$He is predominantly emitted at earlier stage.

\begin{table}[h]
\centering\caption{\label{tab:table1}Energy peak position $E_{\rm p}$ of t and $^3$He for SSDT 2 to 4.}
\renewcommand{\arraystretch}{1.1}
\renewcommand{\tabcolsep}{0.7pc}
\begin{tabular}{cccc}
\hline
  \hline
     & SSDT2 &  SSDT3 & SSDT4    \\
  \hline
   $E_{\rm p}$ of t (MeV)        & 45 &   40 &   19     \\
   $E_{\rm p}$ of $^{3}$He (MeV) & 62 &   58 &   38     \\
 \hline
  \hline
\end{tabular}
\end{table}
 
The energy spectra calculated by ImQMD05 are presented in Fig. \ref{energy_spectra} with solid and dash lines for t and $^3$He, respectively. It should be clarified that the yield of clusters is usually underestimated by transport model. In order to gain a clear view and direct comparison of the shape, the yields of the model predictions are scaled according to the experimental yields, and the correction factor is the same for triton and  $^3$He in a given angular window. It can be seen that the trend of the spectra of t and $^3$He is qualitatively repeated by the model calculations. Switching from t to $^3$He, the energy spectra become slightly harder, and the energy peak positions  move to the right side. At large angles, as shown in panel (b) and (c), the simulated descending tails of  $^3$He spectra agree better with the data compared to that of triton, suggesting that the  high-energy $^3$He is dominated by dynamic emissions. Quantitatively speaking, however, the splitting  between t and $^3$He in model calculations is less pronounced than in the experimental data, particularly at smaller angles. It suggests that the origin and the formation of light clusters, as of t and $^3$He, is  seemingly more complicated than the coalescence approach usually adopted by current transport models.
 We note here that in the ImQMD05 calculations, the fission events are hard to simulate because of its long time scale, so the events are selected by two LCPs firing in SSDTs, $M_{\rm lcp}  \ge  2$. By inspecting the  experimental spectra under the same condition, as shown in the inset of panel (c) for example, the conclusions are not changed.

Benefiting from the wide angular coverage of the SSDTs in laboratory reference frame, the angular distribution of the isobaric yield ratio of $\rm t/^3He$ can be analyzed. To avoid the influence of the possible experimental distortion caused by the  energy threshold in each SSDTs, we focus on the descending part on the right side of the energy peak. The energy peak positions ($E_{\rm p}$) are listed in Tab. \ref{tab:table1}. Meanwhile, using the energy condition $E \ge E_{\rm p}$  as the low limit cut, one can suppress the interference of the evaporation process and emphasize  the feature of the dynamic emission.    


\begin{figure}[h]
 \centering
 \includegraphics[angle=0,width=0.45\textwidth]{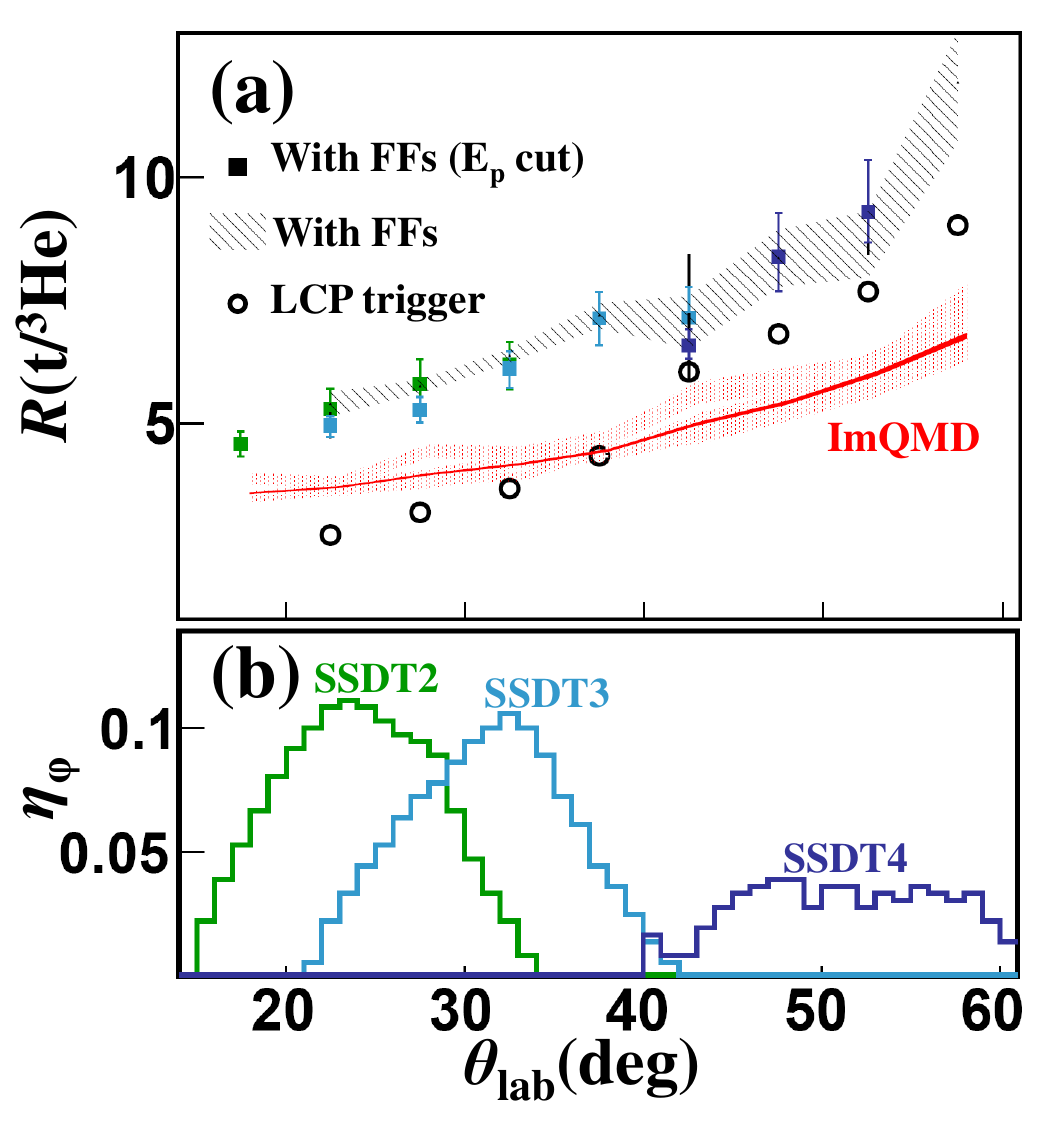}
 \caption{(Color online)  The ratio $R({\rm t/^3He})$ as a function of $\theta_{\rm lab}$ at different conditions. The colored solid squares represent the ratio with $E \ge E_{\rm p}$ cuts in coincidence with fission events, while the diagonal hatched area represents the results by removing the $E_{\rm p}$ cuts. The open circles represents the ratio in the events triggered only by the LCPs in SSDTs with $M_{\rm lcp}  \ge  2$. The red solid curve and the dot-filled band is the ImQMD calculations (see text). (b) The geometric coverage efficiency  $\eta_{\varphi}$ of SSDTs for different laboratory polar angles.}
 \label{ratio}
 \end{figure}

 Fig. \ref{ratio} (a) presents the angular distribution of $R({\rm t/^3He})$ at various conditions as a function of the polar angle in laboratory $\theta_{\rm lab}$, where the geometry efficiency $\eta_{\varphi}$ is plotted in Fig.\ref{ratio}(b).  In panel (a), the colored square markers are the experimental data from the events of one LCP in coincidence with two FFs after the $E_{\rm p}$ cuts are applied. The vertical error bars represent the statistic uncertainties and the colored caps represent the systematic uncertainties arising from  $\pm 1$ MeV variation of the energy $E_{\rm p}$. It is  clearly shown that the angular distribution exhibits a rising trend.
 
 Since $R({\rm t/^3He})$ is positively correlated to the $N/Z$ of emission source \cite{Veselsky:2000kp,Shetty:2003ar,May:2018ulz,Piantelli:2021ybu}, the  behavior of  $R({\rm t/^3He})$ {\it vs.} $\theta_{\rm lab}$ in the wide angular range from $20^\circ$ to $60^\circ$ reveals a convincing picture of the evolution of IDOF. Due to the isospin diffusion and drift mechanisms \cite{Hudan:2014xki}, the neutrons favor to transport from high-density quasi-projectile region to low-density neck region, of which the neutron rich feature has been predicted by various transport model simulations \cite{Sobotka:1994zz,Sobotka:1997zz,Lionti:2005te,Baran:2004ih,Coupland:2011px,Baran:2012zz,zhang2015correlation,Feng:2016hqa,Wang:2018bma}, and experimentally observed in a specific angular window  \cite{Toke:1995zz,Sobotka:2000uj,Shetty:2003ar,Hudan:2005au,Piantelli:2007rs,RodriguezManso:2017emd,INDRA:2017unl,INDRA:2022amz} or a parallel velocity  window \cite{Plagnol:2000fx,Piantelli:2001zb,Shetty:2003ar,Hudan:2005au,Theriault:2006dm,ISOSPIN:2008mur,Kohley:2011zza,Piantelli:2021ybu,Ogul:2023rmp}. 
 
 The robustness of the increasing trend of $R({\rm t/^3He})$ as a function of $\theta_{\rm lab}$ can be further checked by varying the conditions.  If one takes the detector threshold as the low limit cut instead of the $E_{\rm p}$ cut, i.e., $v_{\rm t} \ge 0.09c$ and  $v_{\rm ^3He}\ge 0.15c$, same as in \cite {Wang:2022ysq}, the ratio $R({\rm t/^3He})$  increases similarly, as  plotted by the diagonal hatched band in Fig. \ref {ratio} (a). More interestingly, if one further removes the coincidence with two fission fragments and takes the data triggered by a minimum multiplicity of LCP $M_{\rm lcp}$ firing  the SSDTs (in experiment $M_{\rm lcp} \ge 2$), as shown by the open circles, the increasing trend remains, but the absolute values become markedly smaller. Since SSDT2 has a much higher threshold due to the thicker $\Delta E_1$ and  $\Delta E_2$ than SSDT 3, for the ratios without $E_{\rm p}$ cuts, only the latter is counted. 
 
 Indeed, the enhancement shown by the hatched band compared to the open circles in  Fig. \ref {ratio} (a) indicates that the neck emission is related to the impact parameter. In our experiment, the linear momentum transfer (LMT) derived from the folding angle of the two fission fragments  is averaged at $\rm \left<LMT \right> =0.35$. From a simple geometry picture, it corresponds to an average impact parameter $\left<b \right> \approx 7$ fm, which is a semi-peripherial reactions. On the other hand, the inclusive events triggered by $M_{\rm lcp}  \ge  2$ have a much weaker  $b$ dependence. Thus, due to the overlap of the outer neutron skin of the projectile and the target, the neutron richness of the neck zone formed in the semi-peripherial events tagged by the fission fragments (diagonal hatched band) is more pronounced than in the LCP triggered events (open circles). It has been pointed out that,  the neutron rich neck zone is formed from dynamical effects in a semi-peripherial reactions \cite{Baran:2001pz,Lionti:2005te,RodriguezManso:2017emd}. The difference between the results with and without fission tagging suggests that the neck emission occurs before the isospin equilibrium is fully achieved.

 The rising trend of $R({\rm t/^3He})$ is also qualitatively reproduced by transport model calculations using ImQMD05, as shown by the solid curve with dot-filled area in Fig. \ref{ratio} (a). Here, the dot-filled area represents  the results with $b$ varying from 1.0 to 7.0 fm, and the solid curve is the $b$-averaged value. The parameter of $E_{\rm sym}(\rho)$ is $\gamma=0.5$.  The trigger and cut condition is the same as for the open circles. The calculations confirm the increasing trend but underestimate the increasing rate compared to the data points. Again, the failure of quantitative reproduction of the data suggests  the complication of the clustering mechanism in heavy ion collisions at Fermi energies.  
 
 \begin{figure}[h] 
 \centering
 \includegraphics[angle=0,width=0.98\textwidth]{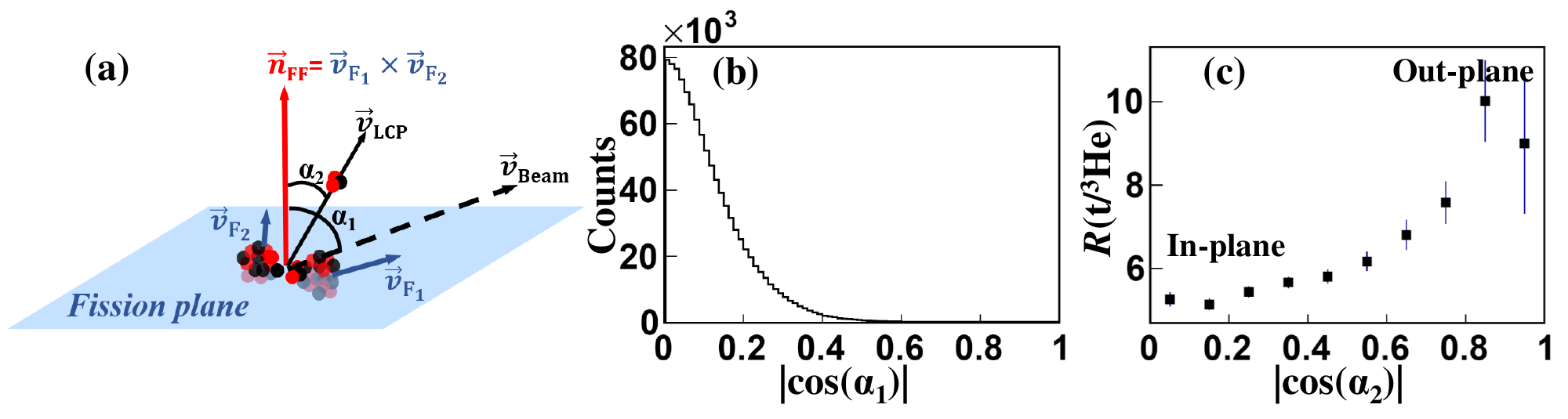}
 \caption{(Color online)  (a) Geometric diagram of fission plane of FFs and neck emission. (b) Angular distribution between the normal vector $\vec{n}_{\rm FF}$ of the fission plane and the beam direction $\vec{v}_{\rm beam}$. (c) Angular distribution between the normal vector of the fission plane and the velocity of LCPs $\vec{v}_{\rm LCP}$. }
 \label{LCP_FFs}
 \end{figure}
 
Finally, the neutron rich neck emission can be confirmed via the ratio $R({\rm t/^3He})$ in and out of the fission plane, as shown in Fig. \ref{LCP_FFs} (a). The fission plane is reconstructed by the velocity of two FFs, using  $\vec{n}_{\rm FF} = \left(\vec{v}_{\rm F_1} \times \vec{v}_{\rm F_2}\right)/|\vec{v}_{\rm F_1}||\vec{v}_{\rm F_2}|$ to denote the normal vector of the plane. Defining  $\alpha_{1}$ as the angle between  $\vec{n}_{\rm FF}$ and the beam direction $\vec{v}_{\rm beam}$, the distribution of $\alpha_{1}$ characterizes how much the fission plane deviates from the beam. As shown in Fig. \ref{LCP_FFs}(b), the $|\rm cos(\alpha_{1})|$ is peaked at 0 with a rather small width $\sigma_{\rm \alpha_1} \approx 6^\circ$, inferring that the fission plane keeps approximately the memory of the initial angular momentum of the rotating system.  

With the above inference, Fig. \ref{LCP_FFs}(c) presents the angular distribution of $R({\rm t/^3He})$ with respect to the fission plane. The  $\alpha_{2}$ on the abscissa is the relative angle between $\vec{n}_{\rm FF}$ and the velocity of the coincident triton or $^3$He $\vec{v}_{\rm LCP}$, with $|\cos (\alpha_{2})|=0$ (1) corresponding to in-plane (out-plane) emission. The $E_{\rm p}$ cuts are applied same as in Fig. \ref{ratio} to calculate the $R({\rm t/^3He})$. The increasing trend of $R({\rm t/^3He})$ with  $|\cos (\alpha_{2})|$ indicates that the out-plane emission is enhanced. In a naive picture, the particles emitted from the neck region favor more out-plane at $\alpha_{2} \approx 0^\circ$ or  $\cos(\alpha_{2}) \approx 1$ because the emission in the fission plane can be blocked by the two heavy fragments. On the other hand, the particles emitted from  both bodies of the dinuclear system, followed by the rupture to quasi-projectile and quasi-target, respectively, tend to populate in-plane at $\alpha_{2} \approx 90^\circ$ because of the centrifugal motion. According to the picture, the increasing trend of $R({\rm t/^3He})$  as a function of $|\cos (\alpha_{2})|$  consistently supports the formation of the neutron rich neck in the reactions. 

 \section  {Summary } 
 The energy spectra and the angular distribution  of triton and $^3$He ranging from $20^\circ$ to $60^\circ$ in the laboratory in coincidence with fission fragments are analyzed in 25 MeV/u $^{86}$Kr +$^{208}$Pb reactions. It is shown that the energy spectra of $^3$He are generally harder than triton, and the effect is more pronounced at small angles. The yield ratio  $R\rm {(t/^3He)}$ exhibits an increasing trend with $\theta_{\rm lab}$, evidencing the evolution of the isospin and density gradient from quasi-projectile to the  mid-velocity neck region. The ImQMD simulations achieve a qualitative description of the energy spectra and angular distribution of t and $^3$He, supporting the dynamic feature of the emission of t and $^3$He from the neck. However, the failure of quantitative reproduction, which can not be cured by the filter conditions in the analysis, indicates a more complicated clustering mechanism than the current  coalescence picture in the transport model and calls for further systematic investigations.  Moreover, the ratio $R\rm {(t/^3He)}$ increases with the angle to the fission plane, in accordance with  the enhancement of the out-plane emission of neutron-rich particles from the neck. The distribution properties of  triton and $^3$He  provide an effective means to understand isospin dynamics related to the neck formation and a fine probe to constrain the $E_{\rm sym}(\rho)$ in heavy ion reactions.

\section { Acknowledgement }

  This work is supported  by the Ministry of Science and Technology of China under Nos.  2022YFE0103400 and 2020YFE0202001, 
  and by the National Natural Science Foundation of China under Grant Nos. 12205160, 
  11961131010, 11961141004, 
  and 11965004, and by the Polish National Science Center under No. 2018/30/Q/ ST2/00185. 
  Z.G.X  is also supported by Tsinghua University Initiative Scientific Research Program and the Heavy Ion Research Facility at Lanzhou (HIRFL). The authors thank Huigan Cheng from SCUT, Zhen Zhang from SYSU and Rui Wang from INFN for their valuable discussions.

 \bibliography{refs}


\end{document}